\documentclass[%
 reprint,
superscriptaddress,
showpacs,
 amsmath,amssymb,
 aps,
]{revtex4-1}

\usepackage{graphicx}
\usepackage{color}
\usepackage{verbatim}
\usepackage{subfigure}

\begin{document}

\preprint{APS/123-QED}

\title{Minimum resources for versatile continuous variable entanglement in integrated nonlinear waveguides} 

\author{David Barral} 
\email{Corresponding author: david.barral@c2n.upsaclay.fr}
\author{Kamel Bencheikh}
\affiliation{Centre de Nanosciences et de Nanotechnologies C2N, CNRS, Universit\'e Paris-Saclay, Route de Nozay, 91460 Marcoussis, France}
\author{Virginia D'Auria}
\author{S\'ebastien Tanzilli}
\affiliation{Universit\'e C\^ote d'Azur, Institut de Physique de Nice (INPHYNI), CNRS UMR 7010, Parc Valrose, 06108 Nice Cedex 2, France}
\author{Nadia Belabas}
\author{Juan Ariel Levenson}
\affiliation{Centre de Nanosciences et de Nanotechnologies C2N, CNRS, Universit\'e Paris-Saclay, Route de Nozay, 91460 Marcoussis, France}

\begin{abstract}  In a recent paper [Phys. Rev. A {\bf 96}, 053822 (2017)], we proposed a strategy to generate bipartite and quadripartite continuous-variable entanglement of bright quantum states based on degenerate down-conversion in a pair of evanescently coupled nonlinear $\chi^{(2)}$ waveguides. Here, we show that the resources needed for obtaining these features can be optimized by exploiting the regime of second harmonic generation: the combination of depletion and coupling among pump beams indeed supplies all necessary wavelengths and appropriate phase mismatch along propagation. Our device thus entangles the two fundamental classical input fields without the participation of any harmonic ancilla. Depending on the propagation distance, the generated harmonics are entangled in bright or vacuum modes. We also evidence two-color bipartite and quadripartite entanglement over the interacting modes. The proposed device represents a boost in continuous-variable integrated quantum optics since it enables a broad range of quantum effects in a very simple scheme, which optimizes the resources and can be easily realized with current technology.
\end{abstract}

\maketitle 

\section{Introduction}

\begin{figure}[t]
\centering
\includegraphics[width=0.47\textwidth]{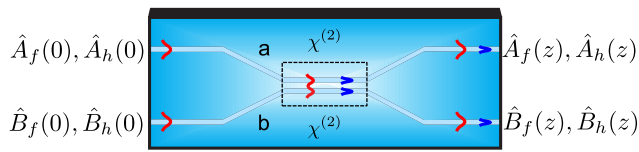}
\vspace {0cm}\,
\hspace{0cm}\caption{\label{F1}\small{(Color online) Sketch of the proposed nonlinear directional coupler made of two identical waveguides a and b with second-order susceptibilities $\chi^{(2)}$. The phase matching condition is only fulfilled in the coupling zone (dashed box). Two input fundamental fields excited in coherent states produce second harmonic fields in this region through SHG. In red are the fundamental waves, evanescently coupled (f). In blue, and more confined, are the non-interacting second harmonic waves generated (h).}}
\end {figure}

The merge between continuous variables (CV) and integrated optics is a hot topic in the quantum optics community \cite{Orieux2016, Kaiser2016, Dutt2015, Stefszky2017, Masada2015, Porto2017}. By exploiting strong optical confinement, integrated nonlinear optics permits one to enhance the efficiency of quantum light sources, in broadband single pass configuration \cite{Herec2003, Barral2017}. In addition, it allows cascading multiple optical functions on a single component, thus ensuring {crucial} benefits in terms of device stability, compactness and loss reduction. Even more attractively, an adequate integration strategy {can strongly diminish} the number of resources that are required to implement a given quantum function. In this article, we apply this concept and show that versatile and simultaneous generation of bi- and quadripartite entanglement is possible by exploiting second harmonic generation (SHG) in coupled nonlinear waveguides. More specifically, we analyze the performances of a nonlinear coupler of the kind of Figure \ref{F1}, designed in such a way that fundamental input beams at a frequency $\omega_{f}$  {are coupled evanescently} and, simultaneously, undergo an SHG process.

As a first result, we demonstrate that the device entangles the input fundamental beams. We note that, compared to spontaneous parametric down conversion (SPDC) or amplification, this strategy {yields} entanglement at a frequency $\omega_{f}$ without {using} ancillary pump beams at higher frequencies \cite{Herec2003, Barral2017}. By exploiting such a property, standard lasers and amplifiers from classical telecommunication technology {could} directly produce entanglement at telecom wavelength without {the need of} auxiliary frequency conversion stages. Afterwards, we demonstrate that {the simultaneous interplay of} fundamental modesÕ up-conversion and coupling also produces entanglement between the noninteracting second harmonic beams at frequency $\omega_{h}=2 \omega_{f}$. Strikingly, by an adequate choice of the coupler geometry, these fields can be set to present features corresponding to Einstein-Podolsky-Rosen (EPR) states (two-mode squeezed vacuum) such as those generated by SPDC \cite{Schori2002}. This scheme opens the possibility to generate entanglement at mid-infrared wavelengths using as resource widely available telecom lasers instead of blue or UV lasers. Eventually, we show the coexistence of two-color bipartite and quatripartite entanglement among fundamental and second harmonic modes.

The ensemble of these remarkable features proves the outstanding capabilities of the nonlinear coupler as versatile and powerful resource {for the flourishing field of CV quantum information} \cite{Cerf2007}. It is to be noted that this device has no bulk-optics analog, as it strongly relies on distributed coupling and nonlinearity that are only accessible to guided-wave nonlinear components. We prove this unique aspect by comparing the performances of our proposed component with those of an integrated two-mode squeezer, where SHG and nonlinear coupling occur in sequence as in bulk-optics schemes for generation of dual-rail two-mode squeezed states \cite{Jin2014, Sansoni2017}.

The article is organized as follows: In section II, we introduce the device under investigation and recall the equations which run the propagation and generation of quantum fields. We then describe the classical and quantum propagation of light in the SHG configuration. In sections III and IV we study the generation and evolution of bipartite and quadripartite entanglement in the device, respectively. In section V we compare the performance of our device with that of the integrated two-mode squeezer. Finally, the main results of this work are summarized in section VI.

\section{SHG in the nonlinear directional coupler} The nonlinear directional coupler, sketched in Figure \ref{F1}, is made of two identical $\chi^{(2)}$ waveguides in which SHG takes place. In each waveguide, two fundamental photons from an input coherent state at frequency $\omega_{f}$ are up-converted into one second-harmonic photon at frequency $\omega_{h}$, all in the same polarization mode. 
We consider that the phase matching condition is fulfilled only in the coupling zone. The energy of the fundamental modes propagating in each waveguide is exchanged between the coupled waveguides through evanescent waves, whereas the interplay of the generated second harmonic waves is negligible for the considered propagation lengths due to their high confinement into the guiding region {\cite{Noda1981}}. The relevant operator which describes this system is the interaction momentum $\hat{M}=\hbar\, \{ g\,\hat{A}_{h} \hat{A}_{f}^{\dag\,2} + g\,\hat{B}_{h} \hat{B}_{f}^{\dag \,2} +C\,\hat{A}_{f} \hat{B}_{f}^{\dag} + H.c.\}$, where $\hat{A}$ and $\hat{B}$ are slowly varying amplitude annihilation operators of fundamental (f) and second harmonic (h) photons corresponding to the upper (a) and lower (b) waveguides, respectively, $g$ is the nonlinear constant proportional to $\chi^{(2)}$, $C$ the linear coupling constant, $\hbar$ the Planck constant, and ${\it H.c.}$ stands for Hermitian conjugate. From this momentum operator, the following Heisenberg equations are obtained \cite{Abram1987, Linares2008}
\begin{align} \label{QS1}
\frac{d \hat{A}_{f}}{d z}=&\, i C \hat{B}_{f} +2 i g \hat{A}_{h} \hat{A}_{f}^{\dag}, \quad \frac{d \hat{A}_{h}}{d z}= i g \hat{A}_{f}^{2}, \nonumber \\
\frac{d \hat{B}_{f}}{d z}=&\, i C \hat{A}_{f} +2 i g \hat{B}_{h} \hat{B}_{f}^{\dag}, \quad \frac{d \hat{B}_{h}}{d z}= i g \hat{B}_{f}^{2},
\end{align}
where $z$ is the coordinate corresponding to the direction of propagation, { and $C$ and $g$ have been taken as real without loss of generality. }

To gain physical insight we linearize and solve the propagation of the quantum states for a specific available technology, although the analysis applies to any material substrate. We consider lithium niobate waveguides. Only the coupling region is periodically poled (PPLN) to compensate for the phase mismatch between the fundamental and harmonic waves and ensure an efficient second-order nonlinear effect. We further consider $C=8 \times 10^{-2}$ \,mm$^{-1}$ and $g=25 \times 10^{-4}$ \,mm$^{-1}$ mW$^{-1/2}$, which will be used in the remainder of the paper. These are standard values in PPLN waveguides \cite{Alibart2016}. 
Unlike SPDC, the undepleted approximation which linearize Eqs.  (\ref{QS1}) can not be used in the SHG case \cite{Herec2003}. We thus implement the linearization of the equations by means of quantum-fluctuation operators $\hat{a}_{j}=\hat{A}_{j}-\alpha_{j}$ and $\hat{b}_{j}=\hat{B}_{j}-\beta_{j}$, with $\alpha_{j}$ and $\beta_{j}$ the mean values related to the input operators $\hat{A}_{j}, \hat{B}_{j}$, with $j=f, h$; a scheme that holds for periodically poled systems \cite{Reynaud1992, Bencheikh1995}. These new operators exhibit zero mean values and the same variances as the input operators. This method was recently used in the analysis of this device in the SPDC and {optical parametric amplification (OPA) regimes} \cite{Barral2017}. In the following we adopt the {same} procedure and normalizations \cite{Note1}. Under the linearization approximation, we first solve the propagation of the classical fields $\alpha_{f}(\alpha_{h})$ and $\beta_{f} (\beta_{h})$ (zeroth order in quantum fluctuations) with appropriate initial conditions $\alpha_{j}(0), \beta_{j}(0)$ ($\alpha_{h}(0)=\beta_{h}(0)=0$ for SHG), to obtain the evolution of the quantum fluctuations. In order to solve the classical equations, we use dimensionless amplitudes  $u_{j} (v_{j})$ and phases $\theta_{j} (\phi_{j})$ related to the {classical fields via $\alpha_{j}=\sqrt{P} \,u_{j} \exp{(i\, \theta_{j})}$, $\beta_{j}=\sqrt{P} \,v_{j} \exp{(i\,\phi_{j})}$, with $P$ the total input energy. We also introduce a normalized propagation coordinate $\zeta=\sqrt{2 P} g z$, which is defined only in the coupling region where phase matching is guaranteed}. {Applying this change of variables into the classical version of Equations (\ref{QS1}), we obtain
\begin{align}\label{us}
\frac{d {u}_{f}}{d \zeta}=&  -\kappa \, {v}_{f} \sin(\phi_{f}-\theta_{f}) - u_{f} u_{h} \sin(\Delta\theta), \nonumber  \\
\frac{d {\theta}_{f}}{d \zeta}=& \kappa \frac{{v}_{f}}{u_{f}} \cos(\phi_{f}-\theta_{f}) + u_{h} \cos(\Delta\theta), \nonumber\\
\frac{d {u}_{h}}{d \zeta}=& u_{f}^2 \sin(\Delta\theta), \qquad \frac{d {\theta}_{h}}{d \zeta}= \frac{u_{f}^2}{u_{h}} \cos(\Delta\theta).
\end{align}
Four additional equations can be obtained by exchanging $u\leftrightarrow v$ and $\theta \leftrightarrow \phi$.} The two governing parameters of the system are the nonlinear phase mismatch $\Delta\theta \equiv \theta_{h}-2\theta_{f}$ $(\Delta\phi \equiv \phi_{h}-2\phi_{f})$ and the effective coupling $\kappa=C/(\sqrt{2P} g)$. The nonlinear phase mismatch drives the nonlinear optical processes whereas the effective coupling indicates which of the two competing effects is stronger, either the linear or the nonlinear interactions.




 \begin{figure}[t]
 \centering
 \includegraphics[width=0.47\textwidth]{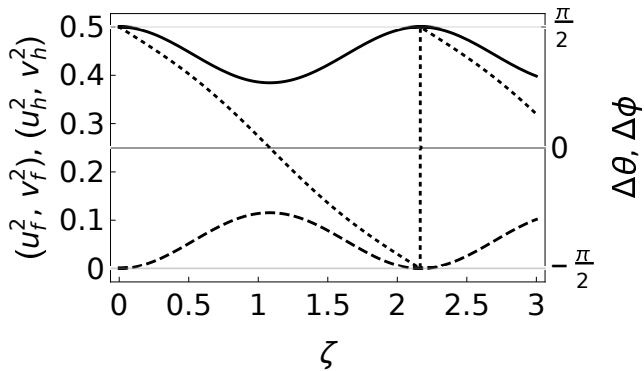}
\vspace {-0.25cm}\,
\hspace{0cm}\caption{\label{F2}\small{Classical-fields power and phase difference propagation. Dimensionless fundamental (solid) and second harmonic (dash) powers, and phase difference of the classical waves $\Delta\theta (\Delta\phi)$ (dot). $\kappa=1.13$ and $P_{h}/ P_{f}=10^{-18}$. $\zeta$ is the normalized propagation coordinate.}}
\end {figure}

In the SHG regime, the classical operation of {the nonlinear directional coupler} as an all-optical switch was {numerically analyzed} in Ref. \cite{Assanto1993}; however, propagation was not dealt with in that work. {The spectral quantum correlations produced in a nonlinear directional coupler inside a Fabry-Perot cavity were theoretically evaluated in Ref. \cite{Mallon2008}. However, in that work the specifics of cavities were used, i.e., steady-state solutions and coupling between the harmonic fields, whereas our approach deals with single-pass traveling waves and noninteracting harmonic fields.} Since there is no known exact analytical solution to Equations (\ref{us}), we solve them numerically. We set fundamental input powers and phases to be equal in each waveguide, which leads to the excitation of the even fundamental supermode related to the coupled system \cite{Yariv1988}. {Inasmuch as numerical simulations cannot deal with input vacuum states, we input harmonic coherent states with a mean number of photons very close to zero ($< 1$ photon). We thus set the ratio between the fundamental and harmonic powers at each waveguide as $P_{h}/ P_{f}=10^{-18}$, such that $u_{f}^{2}(0)=v_{f}^{2}(0)\approx 1/2$ and $u_{h}^{2}(0)=v_{h}^{2}(0) \approx 0$.} The harmonic initial phases are set equal as those of fundamentals. {It should be noted that in SPDC and OPA $\kappa=1$ represents the threshold for which the linearization approximation remains valid \cite{Barral2017}. In SHG, however, the regime of the quantum-noise-induced fundamental mode revivals appears only when $\kappa\rightarrow 0$, leading to an exponential growth of the fluctuations and the breakdown of the linear approach \cite{Olsen2000, Mallon2008}. Below we choose an effective coupling $\kappa=1.13$, since it is a feasible value with PPLN technology and for the sake of comparison  with the results obtained for SPDC and OPA in ref. \cite{Barral2017}. $\zeta=1$ then stands for an interaction length $z\approx 14$ mm, a coupling length accessible with present fabrication technology. We emphasize, however, that our approach remains valid for values of $\kappa$ as small as $0.02$.}

{Figure \ref{F2} displays the dimensionless classical powers for each mode in each waveguide and the nonlinear phase mismatch along the propagation.} Although all parameters other than the initial conditions are identical to those in Ref. \cite{Barral2017}, where  $u_{f}^{2}(0)=v_{f}^{2}(0)\approx 0$ and $u_{h}^{2}(0)=v_{h}^{2}(0) \approx 1/2$, a strong fundamental field depletion is observed here in SHG in contrast with the harmonic undepletion in SPDC (Figure 2a in Ref. \cite{Barral2017}).
{In both cases the classical fundamental and harmonic powers are {solely} driven by the coupling-based nonlinear phase mismatch $\Delta\theta\,(\Delta \phi)$, {since the single phases have the same evolution along propagation $[\theta_{f}(\zeta)=\phi_{f}(\zeta)]$}. The linear coupling of the fundamental modes yields this phase mismatch, which cyclically destroys the phase matching produced by the PPLN, driving two cascaded nonlinear optical processes: up-conversion followed by down-conversion. Thus the evolving phase mismatch periodically switches the system from an efficient fundamental-to-harmonic conversion to an efficient harmonic-to-fundamental conversion. In the SHG case, as soon as the propagation in the waveguides starts, this phase difference jumps to $\pi/2$ and evolves down to $-\pi/2$ cyclically. Note that this phase jump and evolution also arises in SHG with a depleted input in uncoupled single waveguides with imperfect phase matching \cite{Armstrong1962}.}

The solutions of the classical system of equations are then fed into first-order equations in the quantum fluctuations keeping only the linear terms. We solve the evolution of the amplitude and phase quadratures of the field related to each optical mode, $\hat{X}^{(A,B)}_{(f,h)}$ and $\hat{Y}^{(A,B)}_{(f,h)}$ \cite{Barral2017}. {In terms of dimensionless variables, the propagation of the quantum field quadratures are given by  \cite{Li1994}
\begin{align} \label{Quad1}
\frac{d \hat{X}_{f}^{A}}{d \zeta}= &- u_{h} \sin(\theta_{h}) \hat{X}_{f}^{A} + u_{h} \cos(\theta_{h}) \hat{Y}_{f}^{A}  - \kappa \hat{Y}_{f}^{B} \nonumber \\ 
&+ \sqrt{2} u_{f} \sin (\theta_{f}) \hat{X}_{h}^{A} - \sqrt{2} u_{f} \cos (\theta_{f}) \hat{Y}_{h}^{A},   \nonumber  \\ 
\frac{d \hat{Y}_{f}^{A}}{d \zeta}=  &\, u_{h} \cos(\theta_{h}) \hat{X}_{f}^{A} + u_{h} \sin(\theta_{h}) \hat{Y}_{f}^{A}  + \kappa \hat{X}_{h}^{B}  \nonumber \\ 
&+ \sqrt{2} u_{f} \cos (\theta_{f}) \hat{X}_{h}^{A} + \sqrt{2} u_{f} \sin (\theta_{f}) \hat{Y}_{h}^{A},    \nonumber   \\  
\frac{d \hat{X}_{h}^{A}}{d \zeta}=    &-\sqrt{2} u_{f} \sin(\theta_{f}) \hat{X}_{f}^{A} - \sqrt{2} u_{f} \cos(\theta_{f}) \hat{Y}_{f}^{A},   \nonumber \\
\frac{d \hat{Y}_{h}^{A}}{d \zeta}=  &\, \sqrt{2} u_{f} \cos(\theta_{f}) \hat{X}_{f}^{A} - \sqrt{2} u_{f} \sin(\theta_{f}) \hat{Y}_{f}^{A},
\end{align}
and the other four equations are obtained by exchanging again $u\leftrightarrow v$, $\theta \leftrightarrow \phi$ and $A \leftrightarrow B$. This system of equations can be rewritten in compact form as $d \hat{\xi}/ d \zeta = \mathbf{\Delta}(\zeta)\, \hat{\xi}$, where $\mathbf{\Delta}(\zeta)$ is a $8\times8$ matrix of coefficients, and $\hat{\xi}=(\hat{X}_{f}^{A},\hat{Y}_{f}^{A},\hat{X}_{h}^{A},\hat{Y}_{h}^{A},\hat{X}_{f}^{B},\hat{Y}_{f}^{B},\hat{X}_{h}^{B},\hat{Y}_{h}^{B})^T$. The formal solution of this equation is given by $\hat{\xi}(\zeta)=\mathbf{S}(\zeta)\, \hat{\xi}(0)$, with the evolution operator $\mathbf{S}(\zeta)=\exp\{\int_{0}^{\zeta}\mathbf{\Delta}(\zeta')\, d \zeta' \}$ \cite{Herec2003}. Experimentally, the most interesting observables of our system in terms of CV entanglement are the second-order moments of the quadrature operators, elements of the covariance matrix $\mathbf{V}$: $V({\xi}_{j}^{O}, {\xi}_{k}^{O'})=\frac{1}{2}(<\Delta\hat{\xi}_{j}^{O} \Delta\hat{\xi}_{k}^{O'}> + <\Delta\hat{\xi}_{k}^{O'} \Delta\hat{\xi}_{j}^{O}>)$, with $\Delta \hat{\xi}\equiv\hat{\xi}-\langle\hat{\xi} \rangle$, and where $i,j=f,h$, and $O,O'=A,B$ \cite{Adesso2014}. $\mathbf{V}$ is a real symmetric matrix that contains all the useful information about the quantum states propagating in the device, and $\mathbf{V}$ can be efficiently measured by means of homodyne detection \cite{Dauria2009} or quasiresonant analysis cavities in the case of bright beams \cite{Coelho2009}. The covariance matrix at any normalized propagation plane $\zeta$ is given by $\mathbf{V}(\zeta)=\mathbf{S}(\zeta)\, \mathbf{V(0)} \,\mathbf{S}^{T}(\zeta)$, where $\mathbf{V(0)}=(1/2)\, \mathbf{1}$ is the covariance matrix related to the input Gaussian fields, with a 1/2 shot noise in our convention.} Evolution of squeezing and quantum correlations between any pair of quadratures can be obtained at any length from the elements of this matrix.


 \begin{figure}[t]
  \centering
\includegraphics[width=0.45\textwidth]{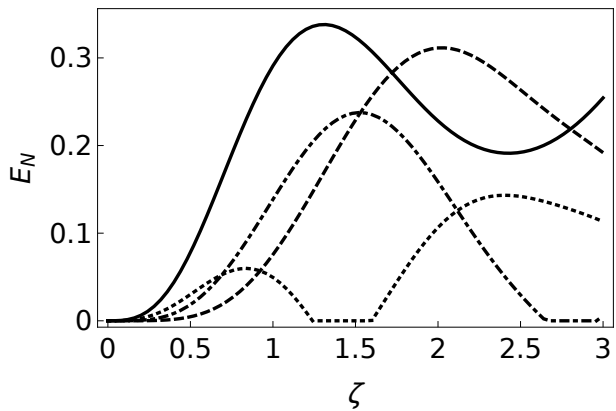}
\vspace {0cm}\,
\hspace{0cm}\caption{\label{F3}\small{Bipartite entanglement between single-mode parties. Logarithmic negativity $E_{\mathcal{N}}$ corresponding to the subsystem of fundamentals (solid), harmonics (dash), fundamental-harmonic in the same waveguide (dot), and fundamental-harmonic in different waveguides (dot-dash). $\kappa=1.13$ and $P_{h}/ P_{f}=10^{-18}$. $\zeta$ is the normalized propagation coordinate.}}
\end {figure}

\section{Bipartite entanglement} The amount of CV entanglement in bipartite splittings of the system is easily quantified through the logarithmic negativity $E_{\mathcal{N}}$ {\cite{Vidal2002, Plenio2005}}. It can be obtained from the covariance matrix $\mathbf{V}$ and is defined in such a way that any value $E_{\mathcal{N}}>0$ indicates entanglement in any bipartite splitting of the system, made up of one or various modes. Figure \ref{F3} shows the $E_{\mathcal{N}}$ corresponding to two modes of the quadripartite system, i.e., tracing out the two modes not considered. Notably, both the fundamental modes (solid) and the generated noninteracting second harmonic modes (dash) are entangled for all the propagation distances here considered, whereas entanglement between the fundamental and harmonic modes propagating in the same waveguide (dot) and in different waveguides (dash-dot) is obtained at specific distances. The main causes of these effects are the simultaneous depletion and evanescent coupling of the fundamental modes and the periodic alternation between up- and down-conversion driven by the coupling-based nonlinear phase mismatch $\Delta\theta (\Delta\phi)$. 

In more detail, the entanglement of the fundamental modes is caused by depletion-based squeezing and linear coupling of the input coherent fields. This is similar to the entanglement of two single-mode squeezed states in a bulk-optics beam splitter \cite{Kim2002}, but in a distributed way. Further comparison is given at the final section. In the case of the harmonic modes the supermodes framework enables a clearer view: as the even fundamental supermode propagates along the nonlinear directional coupler, a pair of supermode photons is up-converted to an harmonic photon which is delocalized in the two waveguides, like an effective harmonic supermode. Therefore the harmonic fields are entangled in the individual modes basis. The device acts thus as a distributed nonlinear beam splitter, where two input fundamental photons propagating in each waveguide are transformed in one entangled dual-rail harmonic photon. Both fundamental modes and harmonic modes present maximum values of entanglement for device lengths less than $z\approx 3.5$ cm, feasible with current technology. Strikingly, at $\zeta_{0} \approx 2.1$ the harmonic fields present zero mean value (Figure \ref{F2}) and a value of entanglement $E_{\mathcal{N}}\approx 1/3$. Analyzing the entries in the main diagonal of the covariance matrix $\mathbf{V}$, squeezing in the harmonic modes is also found (not shown). These features and the Gaussian nature of the states under study lead us to conclude that an EPR state, or two-mode squeezed vacuum, is present in the harmonic modes at this propagation distance. {It is important to emphasize that, unlike the total system that is a pure state,  each single-color subsystem is a mixed state. The purity of the quantum state corresponding to the harmonic modes can be obtained from the covariance matrix related to that subsystem, $\mathbf{V}_{h}$, through $\mu_{h}=1/(4\sqrt{\det \mathbf{V}_{h}})$ \cite{Adesso2014}. In this case a purity $\mu_{h}=97.4 \%$ is found at $\zeta_{0}$. Likewise, the fidelity of this harmonic mixed state with regard to a two-mode squeezed vacuum of covariance matrix $\mathbf{V}_{sq}$ and squeezing parameter $r$ can also be easily worked out from $\mathbf{V}_{h}$ \cite{Spedalieri2013}. In this case it is given by $\mathcal{F}\approx1/\sqrt{\det (\mathbf{V}_{h}+\mathbf{V}_{sq})}$. At $\zeta_{0}$, a value of $\mathcal{F}=98.0\%$ is obtained for $r=0.11$, equivalent to -1 dB squeezing.} Further, fundamental-modes depletion leads to two-color entanglement between the modes propagating in each waveguide and in different waveguides through linear coupling of the fundamental fields \cite{Li1994, Olsen2004}.

The above features present important applications at the technological level. First, compared to standard SPDC, squeezing at frequency $\omega_{f}$ is obtained at the output of the nonlinear directional coupler without the need for additional frequency doubling stages to generate pump beams. One can thus use the same laser in both generation and detection stages, simplifying setups and avoiding problems of mode matching. This simple device opens the possibility of producing bright entangled states on demand at telecom wavelengths, where low-loss optical fibers and high-performance standard components are available. On top of this, the generation of EPR states at the harmonic frequency is an asset. It could represent eventually a novel way of generating twin photons, but further investigation in this direction has to be carried out. Notably, there are distances where values of entanglement as high as $E_{\mathcal{N}}\approx 1/3$ are found in both fundamental and harmonic subsystems. Bipartite entanglement increases as $\kappa$ decreases.  When doubling the total input power ($\kappa=0.8$) peaks of $E_{\mathcal{N}}\approx 2/3$ in both single-mode parties are obtained. These values are on the order of those reported with the nonlinear directional coupler in an OPA regime \cite{Barral2017} or with schemes involving optical cavities \cite{Mallon2008}. These bipartite entangled states are the resources of prominent CV-based quantum protocols such as quantum teleportation \cite{Ralph1998}, quantum cryptography \cite{Bencheikh2001}, quantum imaging \cite{Boyer2008}, and optomechanical entanglement \cite{Mazzola2011}.

\begin{figure}[t]
\centering
\includegraphics[width=0.45\textwidth]{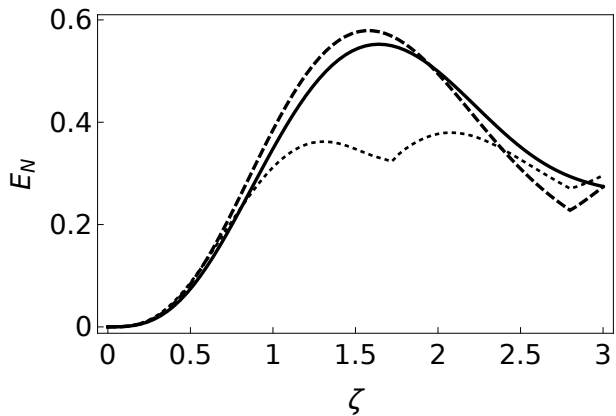}
\vspace {0cm}\,
\hspace{0cm}\caption{\label{F4}\small{Bipartite entanglement between two-mode parties. Logarithmic negativity $E_{\mathcal{N}}$ corresponding to the subsystem of (fundamental a, fundamental b)-(harmonic a, harmonic b) (solid), (fundamental a, harmonic a)-(fundamental b, harmonic b) (dash), and (fundamental a, harmonic b)-(fundamental b, harmonic a) (dot). $\kappa=1.13$ and $P_{h}/ P_{f}=10^{-18}$. $\zeta$ is the normalized propagation coordinate.}}
\end {figure}

Bipartite entanglement can also be analyzed when more than two modes are involved. As an example, Figure \ref{F4} shows the logarithmic negativity $E_{\mathcal{N}}$ corresponding to the four modes of the system in a bipartite splitting, i.e., the entanglement of two subsystems made up of two modes each. The subsystems (fundamental a, fundamental b)-(harmonic a, harmonic b) (solid), (fundamental a, harmonic a)-(fundamental b, harmonic b) (dash), and (fundamental a, harmonic b)-(fundamental b, harmonic a) (dot) are sketched. Note the strength hierarchy between them and the single-mode case, obtaining always higher values of entanglement than that corresponding to the parts involved. This feature appears because negativities can only decrease after tracing out a part of the full system, such that {the same trend would be obtained with a different entanglement estimator} \cite{Vidal2002}. A very interesting consequence of this effect is that the measurement of entanglement on one subsystem, single- or two-color, can be used as a nondemolition measure of entanglement on its complementary. 

\begin{figure}[t]
  \centering
\includegraphics[width=0.45\textwidth]{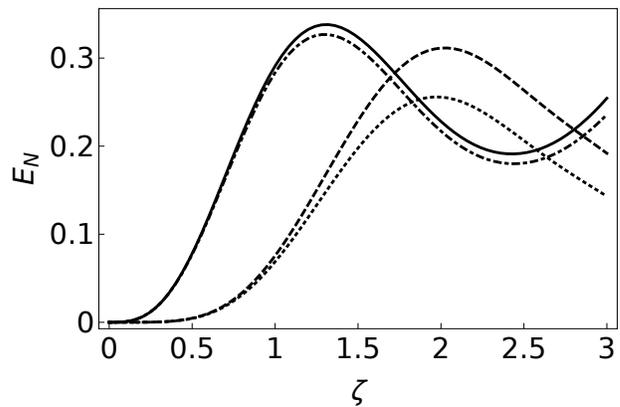}
\vspace {0cm}\,
\hspace{0cm}\caption{\label{F5}\small{{Bipartite entanglement between single-mode parties with propagation losses. Logarithmic negativity $E_{\mathcal{N}}$ corresponding to the subsystem of fundamentals: ideal (solid) and realistic (dot-dash) cases; and harmonics: ideal (dash) and realistic (dot) cases. $\gamma_{f}=0.14$ dB cm$^{-1}$, $\gamma_{h}=0.55$ dB cm$^{-1}$, $\kappa=1.13$ and $P_{h}/ P_{f}=10^{-18}$. $\zeta$ is the normalized propagation coordinate.}}}
\end {figure}

{We can now estimate the influence of losses on the CV entanglement generated in the nonlinear directional coupler. Linear propagation losses $\eta$, such as scattering and absorption, can indeed be easily included in our analysis by inserting fictitious beam splitters with effective transmittivity $\sqrt{\eta}$, mixing appropriately output quantum states with vacuum \cite{Leonhardt1997}. The covariance matrix of the thus computed realistic quantum states $\mathbf{V}^{R}$ is easily found as $V^{R}(\xi_{i}^{O}, \xi_{j}^{O'})=\eta\, V^{I}(\xi_{i}^{O}, \xi_{j}^{O'}) + 1/2 \times (1-\eta)\, \delta_{i,j}\, \delta_{O, O'}$, where $\mathbf{V}^{I}$ is computed from the lossless solutions of Eq. (\ref{Quad1}) and $\delta$ stands for the Kronecker delta. From $\mathbf{V}^{R}$ we can analyze the bipartite entanglement in a nonideal case. Typical values of propagation losses in PPLN waveguides for 780 and 1560 nm are $\gamma_{h}=0.55$ dB cm$^{-1}$ and $\gamma_{f}=0.14$ dB cm$^{-1}$, respectively. These values are included in the covariance matrix by means of $\eta_{i}(\gamma_{i}, z)=e^{-\gamma_{i} z}$. We assume the same losses in both waveguides. Figure \ref{F5} shows the logarithmic negativity $E_{\mathcal{N}}$ corresponding to the  fundamentals and harmonics bipartite splittings of the system deduced from $\mathbf{V}^{I}$ and $\mathbf{V}^{R}$, i.e., for the lossless and lossy cases. A drop in bipartite entanglement of $\approx 3\%$ at $\zeta=1.3$ and $\approx18\%$ at $\zeta=2.1$ is obtained for the fundamental and harmonic fields, respectively. We outline that in both cases the entanglement is quite robust under losses. Note that this analysis can also be extended to extrinsic losses such as coupling, transmission, and detection efficiency. 
}

\section{Quadripartite entanglement} 
\begin{figure}[t]
\centering
\includegraphics[width=0.45\textwidth]{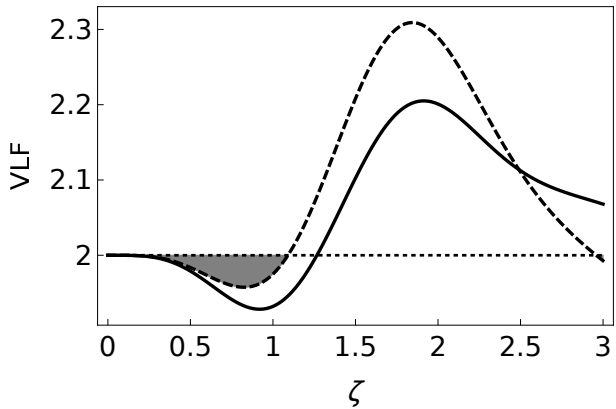}
\vspace {0cm}\,
\hspace{0cm}\caption{\label{F6}\small{Optimized van Loock - Furusawa inequalities (VLF). Simultaneous values under the threshold value imply CV quadripartite entanglement. Solid line: the first and third inequalities. Dash line:  the second inequality. Dot line: quadripartite entanglement threshold. In gray the area where the violation of the three inequalities is obtained. $\kappa=1.13$ and $P_{h}/P_{f}=10^{-18}$. $\zeta$ is the normalized propagation coordinate.}}
\end {figure}

Measuring multipartite full inseparability in CV systems requires the simultaneous fulfillment of a set of conditions which leads to genuine multipartite entanglement when pure states are involved \cite{vanLoock2003, Shalm2013}. This criterion, known as van Loock - Furusawa inequalities, can be easily calculated from the elements of the covariance matrix $\mathbf{V}$ \cite{Migdley2010}. Figure \ref{F6} shows the three inequalities where four arbitrary parameters have been optimized in order to maximize their violation (VLF$<2$). Due to the symmetry of the system, two of the inequalities show equal values (solid). Notably, there are lengths over which all the inequalities are violated, therefore showing two-color quadripartite entanglement within the system (Figure \ref{F6}, gray area). {As for the bipartite entanglement case, a higher degree of entanglement is obtained with lower values of $\kappa$.} Multipartite entanglement can be extended to a higher number of modes by means of waveguide arrays \cite{Rai2012}. These devices could also show multicolor entanglement under appropriate tuning of the parameters. We emphasize that multipartite entanglement of bright beams opens up interesting avenues in CV-based quantum information processing such as multipartite EPR steering \cite{Armstrong2015}.


\section{Comparison with a bulk-optics analog} Finally, let us now compare the performance of the proposed device with an usual bulk-optics scheme for generation of  dual-rail two-mode squeezed states \cite{Kim2002}. To establish a fair comparison, we consider an integrated version of such an approach, that we call the integrated two-mode squeezer.  This is a chip made up of two independent PPLN waveguides which are connected by a linear directional coupler \cite{Jin2014, Sansoni2017, Barral2017b}. Unlike the nonlinear directional coupler, which works in a distributed way, this device operates sequentially: first it produces fundamental-harmonic squeezing and then couples {only the fundamental fields}. The integrated two-mode squeezer can be easily analyzed in the framework of Equations (\ref{QS1}): $C=0$ stands for the uncoupled nonlinear waveguides and $g=0$ for the linear directional coupler.

\begin{figure}[t]
\centering
\hspace{0cm}
\includegraphics[width=0.48\textwidth]{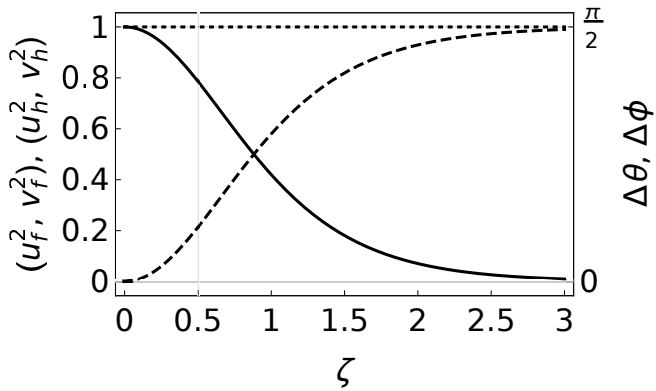}
\vspace {0cm}\,
\hspace{0cm}\caption{\label{F7}\small{Classical-fields power and phase difference propagation in {single nonlinear waveguides}. Dimensionless fundamental (solid) and second harmonic (dash) powers, and phase difference of the classical waves $\Delta\theta (\Delta\phi)$ (dot). $g=25 \times 10^{-4}$ \,mm$^{-1}$ mW$^{-1/2}$ and $P_{h}/ P_{f}=10^{-18}$. $\zeta$ is the normalized propagation coordinate. The vertical line shows the plane equivalent to a physical length of $z=10$ mm.}}
\end {figure}
\begin{figure}[t]
\centering
\hspace{0cm}
\includegraphics[width=0.425\textwidth]{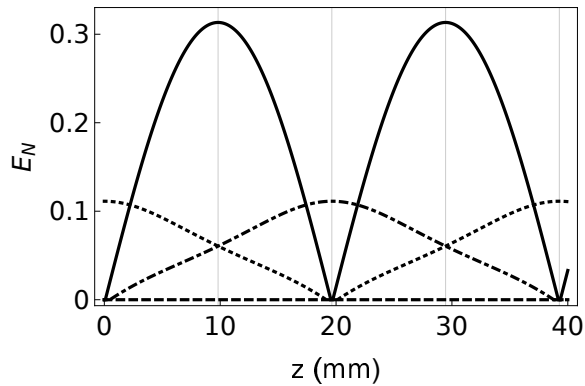}
\vspace {0cm}\,
\hspace{0cm}\caption{\label{F8}\small{Bipartite entanglement in the bulk-optics configuration. Logarithmic negativity $E_{\mathcal{N}}$ corresponding to the subsystem of fundamentals (solid), harmonics (dash), fundamental-harmonic in the same waveguide (dot), and fundamental-harmonic in different waveguides (dot-dash). $C=8 \times 10^{-2}$ \,mm$^{-1}$. The vertical lines show the physical lengths in integer multiples of $L_{ab}/2$.}}
\end {figure}

We analyze first the SHG in each nonlinear waveguide [Equations (\ref{QS1}) with $C=0$]. Figure \ref{F7} shows the dimensionless classical powers for each mode in each nonlinear PPLN waveguide along the propagation. For the sake of comparison, we set the same value of $g$ and the same input power per waveguide as that corresponding to the nonlinear directional coupler. The ratio between the fundamental and harmonic powers is again $P_{h}/ P_{f}=10^{-18}$ and $\zeta=1$ stands for $z\approx 20$ mm. This value is different from that obtained in the nonlinear directional coupler, $z\approx 14$ mm.  {The reason is that in this case the input power $P$ used in $\zeta$ corresponds to that in only one PPLN waveguide, that we set as half of that used previously in order to make a fair comparison between the two configurations}. The fundamental mode experiences a strong depletion fueling the generation of an harmonic wave. As soon as the propagation in each waveguide starts, the phase difference jumps to $\pi/2$, holding this value invariant along propagation due to phase matching \cite{Armstrong1962}. The depletion drives an efficient fundamental-to-harmonic conversion. {It also produces strong single-mode squeezing in both waves and entanglement between them (not shown)} \cite{Lugiato1983, Li1994, Olsen2004}. 
After propagating in the two nonlinear single waveguides, the light finds a linear directional coupler which only couples the fundamental waves [Equations (\ref{QS1}) with $g=0$]. We set the coupling constant with the same value as in the nonlinear directional coupled introduced above. {The length for which the power is totally transferred from one waveguide to the other, the beat length, is here $L_{ab}=\pi/2C\approx19.6$ mm}. Since $g=0$ in this part of the device,  we use as propagation coordinate the actual $z$. Using as input covariance matrix $\mathbf{V}(0)$ the one obtained in the individual nonlinear waveguides, we can calculate its evolution and the entanglement generated following the same steps as above, but without the presence of the nonlinearity. We choose a typical PPLN-waveguides length of 10 mm ($\zeta\approx0.5$) in our simulations \cite{Jin2014}. Figure $\ref{F8}$ shows the logarithmic negativity $E_{\mathcal{N}}$ related to both fundamental and harmonic fields propagating in the linear directional coupler. The fundamental modes (solid) are maximally entangled at $L_{ab}/2$, whereas entanglement disappears at $L_{ab}$. Values as high as $1/3$ are obtained, similar to those obtained with the nonlinear directional coupler (Figure \ref{F3}). Also note that fundamental and harmonic modes propagating in the same waveguide (dot) are entangled at the input of the linear directional coupler, as stated above. The entanglement between different-color fields propagating in the same waveguide (dot) and in different waveguides (dash-dot) is complementary, being maximum (null) for the same (different) waveguides at $2L_{ab}$ and null (maximum) for the different (same) waveguides at $L_{ab}$. However, unlike the nonlinear directional coupler, the harmonic modes are not entangled in this case (dash). As a consequence, quadripartite entanglement is not possible either. 

Therefore, we conclude from the above analysis that the nonlinear directional coupler enables a broader range of effects. Overall, it is also a more compact scheme since the generation and coupling stages are distributed instead of cascaded like in the two-mode squeezer. However, if the aim of the device is just the generation of entanglement between the fundamental waves, the integrated two-mode squeezer could reach higher values than the nonlinear directional coupler at the cost of increasing the PPLN-waveguides length. In the case of the nonlinear directional coupler, an increase in the amount of entanglement could also be obtained by means of suitably engineered periodic PPLN gratings. This structured nonlinear directional coupler would present zones with and without PPLN, whose lengths would be set in order to be always in the up-conversion regime in the PPLN areas, resulting in a larger entanglement between the fundamental fields. The analysis of such structure is beyond the scope of this work.


\section{Conclusion} We have studied the CV entanglement in a nonlinear $\chi^{(2)}$ directional coupler in the second-harmonic generation regime and shown that two input fundamental coherent fields become entangled along propagation due to the combined effect of strong depletion and coupling between them. Remarkably, this effect arises without the need of any ancillary second harmonic field and thus minimizes the resources compared to previous schemes. Waiving the multicolor excitation does not narrow the entanglement capabilities of our minimum-resources device:  we have shown that (i) noninteracting harmonic fields are generated and entangled along the device; (ii) in addition to this bright states entanglemement, a new harmonic two-mode squeezed vacuum arises at specific propagation distances; (iii) measurement of entanglement on any bipartite subsystem, single- or two-color, can moreover be used as a nondemolition measure of entanglement on its complementary as  there are distances where all subsystems exhibit significant values of entanglement; and (iv) two-color quadripartite entanglement is also present in the system under certain conditions. {We have investigated the effect of linear intrinsic losses on entanglement in quantitative terms.} We have also compared the performance of our device, which relies on the distributed combination of coupling and nonlinearity, with the performance of an integrated two-mode squeezer which operates in a sequential way. We found that the nonlinear directional coupler is more compact and gives access to a broader range of effects. Finally, we want to stress that the proposed approach could be relevant for a number of CV quantum protocols. For instance, our integrated platform pumped with telecom C-band wavelength lasers could generate entangled states around 780nm, that could be advantageously interfaced with atomic quantum memories in hybrid quantum protocols \cite{Hacker2016}.

\section*{Acknowledgements}
This work was supported by the Agence Nationale de la Recherche through the INQCA project (Grant Agreement No. PN-II-ID-JRP-RO-FR-2014-0013 and ANR-14-CE26-0038) and the Investissements d'Avenir program (Labex NanoSaclay, Reference ANR-10-LABX-0035).

\end{document}